# Giant magnetic anisotropy of transition-metal dimers on defected graphene

Jun Hu and Ruqian Wu[*]

Department of Physics and Astronomy, University of California, Irvine, California 92697-4575, USA

**Abstract:** Continuous miniaturization of magnetic units in spintronics and quantum computing devices inspires efforts to search for magnetic nanostructures with giant magnetic anisotropy energy (MAE) and high structural stability. Through density functional theory calculations, we found that either Pt-Ir or Os-Ru dimer forms a stable vertical structure on the defected graphene and possess an MAE larger than 60 meV, sufficient for room temperature applications. Interestingly, their MAEs can be conveniently manipulated by using an external electric field, which makes them excellent magnetic units in spintronics and quantum computing devices.

Keywords: giant magnetic anisotropy energy, transition-metal dimer, defected graphene

[*] Corresponding author: wur@uci.edu

Although magnetism is one of the oldest branches of solid-state physics, studies of nanomagnetism are extremely vigorous in recent years, for the use of magnetic units down to the nanometer scale. One of the most challenging problems in this realm is how to frustrate strong thermal fluctuation of magnetization of nanostructures or, in the other word, to increase their blocking temperatures ($T_B$) to above 300 K. The main mechanism that may enhance $T_B$ is the magnetic anisotropy, which measures the energy barrier for flipping the spin moment between two degenerate magnetic states. [1,2,3,4,5] Typical nanostructures including molecular magnets as well as magnetic nanoclusters and nanowires have magnetic anisotropy energies (MAEs) of only a few meV ($T_B$ < 50 K); their magnetic states are hence stable only at very low temperature. [6] For practical applications at room temperature, MAEs of magnetic nanostructures need to be unusually large, up to about 30-50 meV. It was recently reported that several freestanding transition metal dimers such as Rh-Rh, Ir-Ir and Pt-Pt might possess MAEs of 40-70 meV. [7,8,9] However, actual realization of such large MAEs is extremely difficult without a careful search of suitable supporting substrates.

Graphene can be a benign substrate since it is easy to grow and highly stable. Xiao *et al* recently reported giant MAEs of Co-Co and Co-Ir dimers on benzene and graphene, indicating the possibility of making usable magnetic nanostructures on graphene. [10,11] Nonetheless, most 3d adatoms are highly mobile on pristine graphene [12] and various geometrical configurations of Co-Co dimers are almost degenerate; [13] the structural instability of nanostructures on graphene is hence a major concern for room temperature applications. This requires activation of graphene by introducing different defects such as single vacancies (SVs), divacancies (DVs) or nitrogenized divacancies (NDVs) [14,15,16,17]. NDVs with and without insertion of Fe atoms were recently synthesized on carbon nanotubes (CNTs) and the same procedure should be applicable to graphene. [17]

In this paper, we investigate the structural stability and magnetic properties of various transition metal dimers embedded in defected graphene through systematic first-principles calculations. By placing a transition metal dimer on a defect site of graphene in a vertical geometry as depicted in Figure 1a, one of the transition metal atoms (A-type) turns away from graphene so it can retain its magnetization, whereas the other one (B-type) is embedded in the carbon network and hence becomes immobile.

Here, the vacancies are used to anchor atom B, and the vertical geometry is assumed to be the best geometry for the realization of giant MAE. We search for appropriate A and B elements by (1) finding combinations that may have MAEs larger than 30 meV, and (2) testing if the vertical structure for these combinations is stable against swab or drift of A-type atoms. We demonstrated that two of these A-B@defect systems, namely Pt-Ir@SV and Os-Ru@NDV, have giant MAEs of 85 meV and 63 meV as well as high structural stability. With clear understandings from fundamental analyses, we further proposed ways to manipulate MAEs in these systems. For example, the MAE of Pt-Ir@SV can be tuned in a range of 0 meV to 85 meV by applying a moderate electric field.

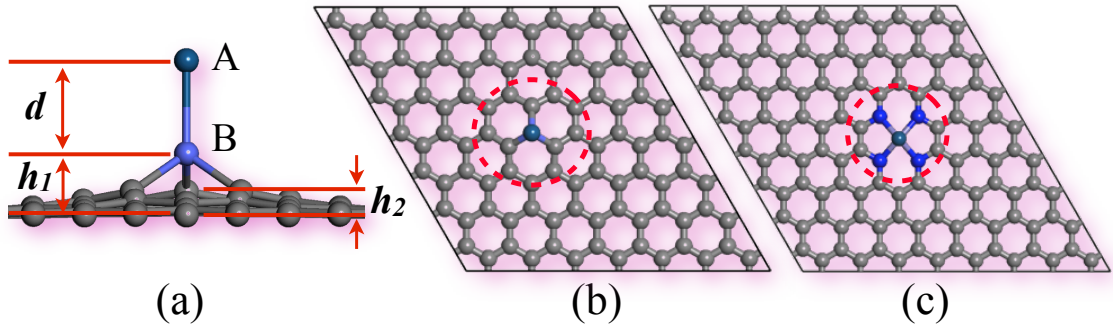

FIG. 1 (color online). (a) Side view of the vertical geometry for a transition metal dimer on defected graphene, denoted as A-B@defect. (b) and (c) Top views of the 7×7 supercell for A-B@SV and the 8×8 supercell for A-B@NDV. The gray and blue spheres stand for C and N atoms, respectively, and the others for TM atoms.

As shown in Figure 1b and 1c, we used either 7×7 or 8×8 supercell along with a 15 Å vacuum between adjacent graphene layers to mimic A-B@SV or A-B@(N)DV cases, respectively. DFT calculations were carried out with the Vienna ab-initio simulation package (VASP), [18,19] at the level of the spin-polarized generalized-gradient approximation (GGA). [20] The interaction between valence electrons and ionic cores was described within the framework of the projector augmented wave (PAW) method. [21,22] The energy cutoff for the plane wave basis expansion was set to 500 eV. A 9×9 k-grid mesh was used to sample the small two-dimensional Brillouin zone. The atomic positions were fully relaxed using the conjugated gradient method for the energy minimization, with a criterion that requires the force on each atom smaller than 0.01 eV/Å. The MAEs were calculated by using the torque method. [23,24,25] Self-consistent calculations with the spin-orbit coupling (SOC) were also performed for selected cases to take the orientation dependence of spin and orbital magnetic moment into account.

The MAE originates from the competition between perpendicular and in-plane contributions of the SOC, which can be expressed approximately in terms of angular momentum operators $L_x$ and $L_z$ (or $L_y$ for magnetic anisotropy in the azimuthal plane) as [26]

$$MAE = \xi^2 \sum_{u,o,\alpha,\beta} (2\delta_{\alpha\beta} - 1) \left[ \frac{|\langle u,\alpha|L_z|o,\beta\rangle|^2}{\varepsilon_{u,\alpha} - \varepsilon_{o,\beta}} - \frac{|\langle u,\alpha|L_x|o,\beta\rangle|^2}{\varepsilon_{u,\alpha} - \varepsilon_{o,\beta}} \right]. \quad (1)$$

Here $\xi$ is the strength of SOC; $\varepsilon_{u,\alpha}$ and $\varepsilon_{o,\beta}$ are the energy levels of the unoccupied states with spin α ($|u,\alpha\rangle$) and occupied states with spin β ($|o,\beta\rangle$), respectively. Obviously, potential candidates for achieving giant MAE should contain heavy elements for the large $\xi$ and, meanwhile, should have narrow d-bands to reduce the denominator in Eq. 1. We considered a series of transition metal dimers, using 5d elements for the A-type atom (i.e., A = Os, Ir or Pt) and Fe group and Co group for the B-type atom (i.e., B = Fe, Ru, Os, Co, Rh or Ir). The Co-Co, Rh-Rh and Ru-Ru dimers were also studied for comparison with the literature. [7,10] The B-type atoms were initially placed at the centers of vacancies in the graphene plane. After relaxation, they move outward by a height of $h_1$ in a range of 1.2 ~ 1.5 Å for A-B@SV or 0.2 ~ 0.5 Å for A-B@NDV as shown in Fig. 1a. This is understandable since SV in graphene is too small to host a metal atom, as was also found in previous studies. [14] On the other hand, the carbon or nitrogen atoms near the metal dimer also shift out of the graphene plane by a height of $h_2$ in a range of 0.4 ~ 0.6 Å for A-B@SV or 0.1 ~ 0.2 Å for A-B@NDV. The distances between two metal atoms, $d$ = 2.2 ~ 2.4 Å, are nonetheless not much affected by the substrate. Significantly, we found that most dimers are unstable on DV, due to the high activity of carbon atoms at the edge of DV. For example, the Os-Co and Ir-Co dimers tilt to the carbon atoms and become nonmagnetic, accompanied by a decrease of total energy of more than 1.0 eV with respect to the vertical geometry. Therefore, nitrogen decoration is necessary to tune the chemical activity of graphene defects toward adsorbates, and we skip results for A-B@DV below.

All vertical A-B@SV and A-B@NDV systems except Ir-Os@NDV and Pt-Os@NDV are magnetic, and the metal atoms couple ferromagnetically. The largest spin moment ($M_S$) of 4.0 $\mu_B$ is found in Os-Fe@NDV and Os-Co@NDV, while the smallest $M_S$ of 0.6 $\mu_B$ occurs in Pt-(Co, Rh or Ir)@SV. It is worthwhile to mention that Os-Fe@NDV and Os-Co@NDV have the same $M_S$, even Co has one more electron than Fe in the 3d shell. We resolved the $M_S$ to different atoms and found that Fe and Os possess 0.91 $\mu_B$ and 2.82 $\mu_B$ in Os-Fe@NDV, whereas Co and Os have 0.62 $\mu_B$ and 3.17 $\mu_B$ in Os-Co@NDV. Clearly, the magnetization of Os is very sensitive to the change of environment and can compensate the difference between Fe and Co. The magnetic moments of transition metal atoms near graphene are significantly quenched, compared to their counterparts in the freestanding cases. For example, $M_S$ of Fe in the free dimers is about 3.0 $\mu_B$, but it decreases to 0.9 ~ 1.4 $\mu_B$ in A-Fe@SV and A-Fe@NDV. More information about the spin moments and spin densities can be found in the Supplemental Material (cf. Table S1).

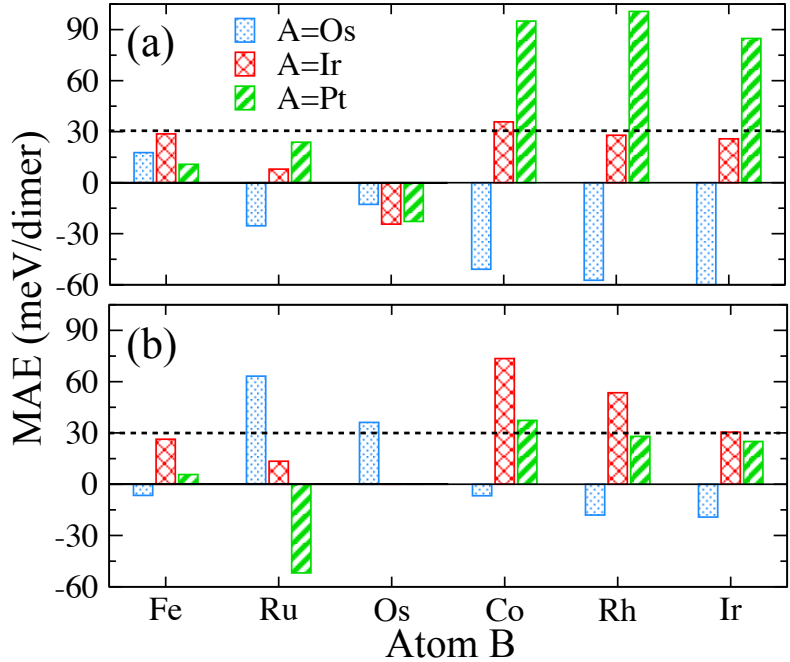

FIG. 2 (color online). Magnetic anisotropy energies of (a) A-B@SV and (b) A-B@NDV. Positive and negative MAEs stand for perpendicular and in-plane easy axes relative to graphene plane, respectively. The horizontal dashed lines mark the position of MAE=30 meV for eye guiding.

Now let us see if giant MAEs may exist in these systems for the use in spintronic devices at room temperature. Note that systems with easy axis perpendicular to the dimer (i.e., MAE<0) are not suitable for applications since their azimuthal MAEs in the horizontal plane are typically very small (e.g. the MAE between x and y directions for Os-Ir@SV is only 0.9 meV). Out of thirty-six A-B@SV and A-B@NDV structures shown in Fig. 2, we found nine cases with positive MAEs larger than 30 meV, namely, Ir-Co@SV, Pt-(Co, Rh, Ir)@SV, Os-(Ru, Os)@NDV, Ir-(Co, Rh)@NDV and Pt-Co@NDV.

For the practical uses, another major concern is regarding their structural stability. There are several mechanisms that may trigger the instability of the vertical geometry of transition metal dimers on graphene. First, the A and B atoms may switch their positions and lose their large MAE. The probability of such position exchange should scale with the energy difference

$$\varDelta E_{pos\text{-}ex} = E\,(B\text{-}A/GR) - E\,(A\text{-}B/GR). \qquad (2)$$

We listed the $\varDelta E_{pos\text{-}ex}$ for all cases with large positive MAEs in table I. Significantly, it appears that Pt strongly prefers the A site, with large positive $\Delta E_{pos\text{-}ex}$ (> 1 eV), whereas Ir and Os tend to switch with other magnetic atoms on both SV and NDV. This is reasonable since both Ir and Os are more active than 3d and 4d elements toward defected graphene, whereas Pt is an "inert" element. It should be pointed out that the position exchange for A-B@NDV is more difficult due to the strong binding between NDV and the B-type atoms, even for cases that have negative $\Delta E_{pos\text{-}ex}$. Therefore, we only rule out Ir-Co@SV at this stage. Note that Ir-Co dimer was predicted to have huge MAE (198 meV) by Xiao et al [11] on pristine graphene, with the Co end touching the carbon ring in the ground state. Our calculations also confirmed their results (MAE=147 meV). However, the energy differences between different adsorption geometries are overall smaller than 0.5 eV so the Ir-Co@graphene structure is unstable at room temperature.

Table I. Position exchange energy ($\Delta E_{pos\text{-}ex}$ in eV) of transition metal atoms. Only the structures of A-B@SV and A-B@NDV with huge MAEs are listed.

|  | A-B@SV | | | | A-B@NDV | | | |
|---|---|---|---|---|---|---|---|---|
| A-B | Ir-Co | Pt-Co | Pt-Rh | Pt-Ir | Os-Ru | Ir-Co | Ir-Rh | Pt-Co |
| $\Delta E_{pos\text{-}ex}$ | -1.15 | 1.39 | 1.23 | 2.80 | -0.57 | -0.96 | -1.18 | 1.26 |

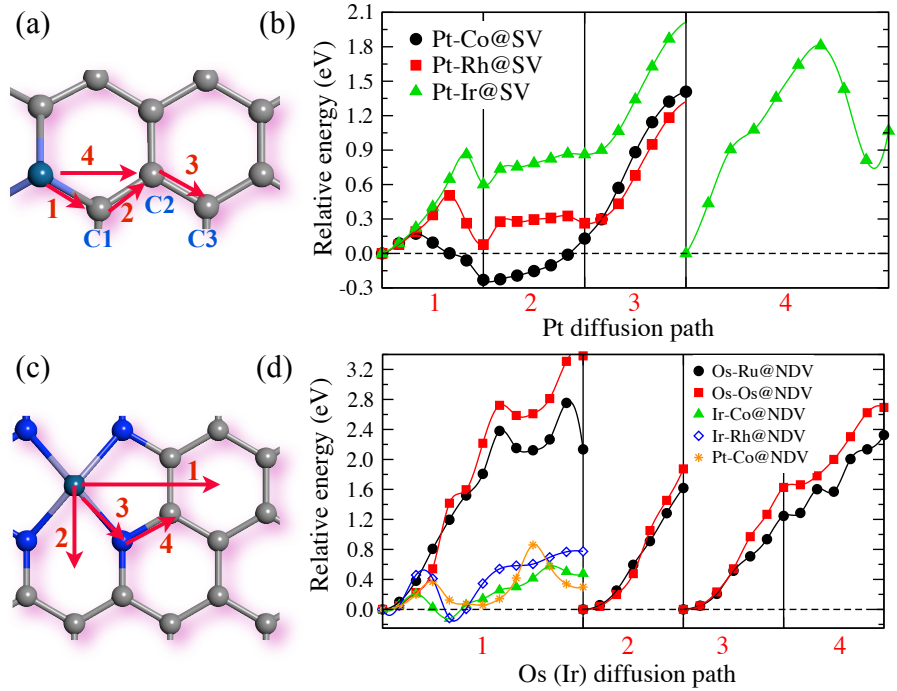

FIG. 3 (color online). (a) and (b) Pt diffusion pathways and energy profiles in Pt-B@SV (B = Co, Rh and Ir). (c) and (d) Ir or Os diffusion pathways and energy profiles in several A-B@NDV structures. For all cases, the total energy of the vertical geometry is set to zero.

The other structural instability associates with the drift of the A-type atoms away from the vertical geometry, depending on the competition between metal-metal bonds and metal-carbon bonds. We considered four possible pathways for the diffusion of Pt atom near SV as sketched in Fig. 3a and the corresponding energy profiles are plotted in Fig. 3b. Pt in Pt-Co@SV tends to take the site above the nearest carbon atom (labeled as C1 in Fig. 3a) and loses its magnetization, indicating stronger Pt-C bond than Pt-Co bond. Note that C1 is activated by the defect and also by Co as shown by the PDOS in Fig. S2, and hence is different from the carbon atom in the flat graphene. Since this process only requires an activation energy of 0.15 eV, we perceive that the vertical geometry for Pt-Co@SV is extremely unstable. Pt-Rh@SV is better, with an energy barrier of about 0.5 eV for Pt segregation, but is still considered unstable at room temperature and its final state also becomes nonmagnetic. Therefore, only Pt-Ir@SV remains as a promising candidate, with an energy barrier of 0.87 eV for Pt diffusion along the path 1 (or 1.8 eV along path 4), a high $\Delta E_{pos-ex}$ of 2.8 eV and a giant positive MAE of 85 meV. Similar analyses for dimers on NDV indicate that Pt-Co, Ir-Co and Ir-Rh cannot keep the vertical geometry since the energy barriers are lower than 0.5 eV for the drift of Pt and Ir atoms to the adjacent bridge sites, as depicted in Fig. 3c and 3d. In contrast, the energy barriers for Os to diffuse away from Ru/Os through all pathways are larger than 1.6 eV. Therefore, Os-Ru@NDV can be another possible candidate for applications, with a giant positive MAE (63 meV) and good structural stability. We want to point out that there is no spin canting for Pt-Ir@SV and Os-Ru@NDV, as their total energies show monotonic angle dependence in Fig. S3(a).

Table II. Binding energies ($E_b$, in eV) for the formation of Pt-Ir@SV and Os-Ru@NDV at different steps: (i) an Ir atom taking SV on graphene: Ir+SV→Ir@SV; (ii) a Pt atom binding to Ir@SV: Pt+Ir@SV→Pt-Ir@SV; (iii) a Ru atom taking NDV on graphene: Ru + NDV → Ru@NDV; (iv) an Os atom binding to Ru@NDV, Os+Ru@NDV→Os-Ru@NDV. For comparison, the cohesive energies ($E_c$, in eV) of bulk Ru, Os, Ir and Pt are also included. For all cases, the energies of the free TM atoms are used as the references.

|       | i    | ii   | iii  | iv   | Ru   | Os   | Ir   | Pt   |
|-------|------|------|------|------|------|------|------|------|
| $E_b$ | 5.29 | 7.89 | 7.31 | 4.29 |      |      |      |      |
| $E_c$ |      |      |      |      | 7.04 | 8.45 | 7.52 | 5.47 |

Although nanostructures are stable as long as there are adequate energy barriers to prevent them from dissociation and aggregation, it is useful to compare the binding of the TM atoms on defected graphene to their bulk phases. As listed in table II, the binding energies for Ir and Ru to defected graphene are as large as 5.29 and 7.31 eV, respectively. The energy gain for attaching the type-A TM atom to a type-B TM atom is also large, i.e., 7.89 eV for Pt-Ir@SV and 4.29 eV for Os-Ru@NDV. Interestingly, the binding energies of Ru to NDV and Pt to Ir@SV are even larger than the cohesive energies of the bulk Ru and Pt, indicating the high stability of these nanostructures. We should point out that even the binding energy of Os to Ru@NDV is smaller than that of the bulk Os (see table II), it is still larger than that of the Os-Os dimer in comparable circumstances (e.g. 4.24 eV for Os to Os@NDV). Moreover, because of the strong attraction from Ru@NDV in a large spatial range, as shown in Fig. 3(d), the chance of forming unwanted Os clusters away from Ru@NDV are not high especially in low temperature.

To understand the driving forces for the development of giant MAEs, we plotted the projected density of states (PDOS) of the $d$ orbitals of both A- and B-type atoms for Pt-Ir@SV and Os-Ru@NDV in Fig. 4. The symmetry around the dimer axis in Pt-Ir@SV is $C_{3v}$ so the $d$ orbitals are split into three groups, with the $d_{xy/x^2-y^2}$ and $d_{xz/yz}$ being two-fold degenerate. In Os-Ru@NDV, the local symmetry reduces to $C_{2v}$, so degeneracies of the $d$ orbitals are lifted. The magnetization of dimers mainly results from the $d_{xy}$ and $d_{x^2-y^2}$ orbitals (see the spin density in Fig. S1). We identified that the couplings between states in the minority spin channel, through $\langle d^u_{xy/x^2-y^2}|L_z|d^o_{xy/x^2-y^2}\rangle$ of Pt and $\langle d^u_{x^2-y^2}|L_z|d^o_{xy}\rangle$ of Os, play the dominant role for the giant positive MAEs of Pr-Ir@SV and Os-Ru@NDV. Although all $5d$ orbitals except $d_{z^2}$ of the A-type atoms (i.e. Pt and Os) are localized, they are well separated in energy and space. The Coulomb correlations between them should be rather moderate and won't substantially affect MAE. As an example, the giant MAE of Os-Ru@NDV maintains in Fig. S3(b) when a reasonable Hubbard U ($\leq$ 2 eV) is added for the d orbitals of Os and Ru.

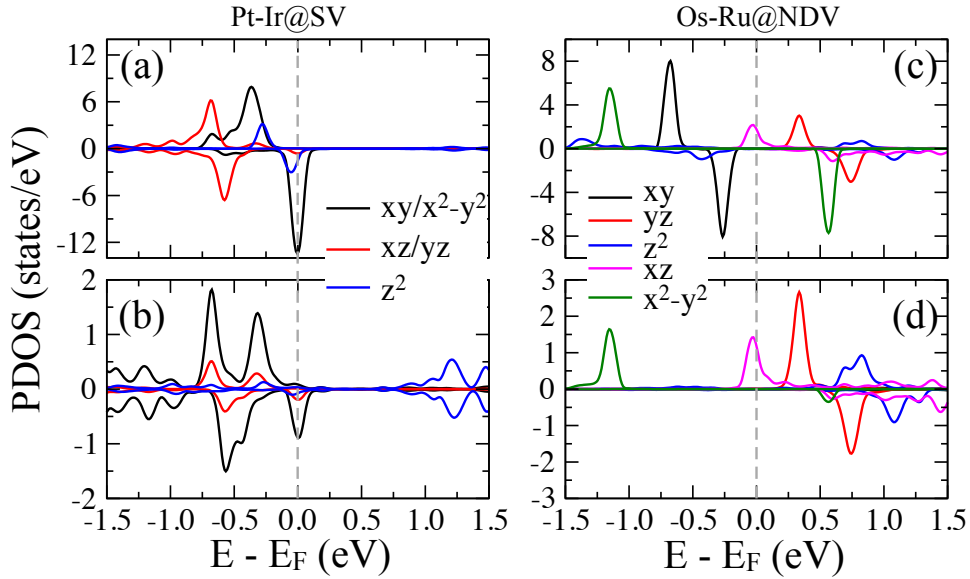

FIG. 4 (color online). Projected density of states (PDOS) of the $d$ orbitals of A-type atoms (upper panels) and B-type atoms (lower panels) of Pt-Ir@SV and Os-Ru@NDV. The vertical dashed lines mark the Fermi level ($E_F$).

To better appreciate and also to manipulate the magnetic anisotropy, it is instructive to plot MAE as a function of the position of the Fermi level according to the rigid band model. As shown in Figs. 5a, the competition between three spin-resolved components, i.e., uu (from states with the same majority spin), dd (from states with the same minority spin), and ud/du (from couplings across two spin channels) produces a narrow peak of MAE near the actual Fermi level, $E_F^0$, for Pt-Ir@SV. This can also be traced to the feature that $Pt - d_{xy/x^2-y^2}$ peak is very narrow and sits right near the Fermi level, as shown in Fig. 4a. In contrast, the plateau of MAE near $E_F^0$ for Os-Ru@NDV appears to be mainly from the minority spin channel (dd), particularly from coupling between the $d_{xy}$ and $d_{x^2-y^2}$ orbitals of Os that show a large energy separation in Fig. 4c.

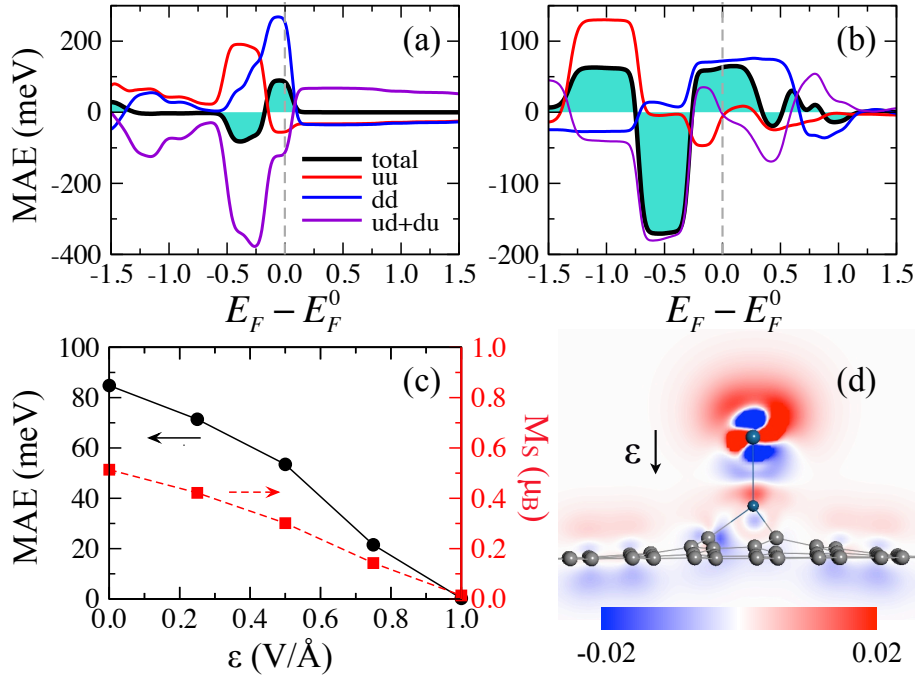

FIG. 5 (color online). (a) and (b) Fermi level dependent total and decomposed MAEs from rigid band model of Pt-Ir@SV and Os-Ru@NDV, respectively. Here uu, dd and ud+du stand for the contributions from SOC interactions between the majority spin states, minority spin states, and cross spin states, respectively. $E_F^0$ stands for the natural Fermi level of the systems. (c) Total MAE and $M_S$ of Pt-Ir@SV under electric field (ε) from 0 to 1.0 V/Å. (d) Electric field induced charge rearrangement, $\Delta\rho = \rho(\varepsilon=1.0) - \rho(\varepsilon=0)$, in a (110) plane that contains Pt and Ir atoms. The arrow shows the direction of the positive electric field.

Interestingly, the total MAE of Pt-Ir@SV rapidly decreases to zero when the $E_F$

shifts upward, which corresponds to more electron occupancy into the $d_{xy/x^2-y^2}$ orbital of Pt. This gives us an easy way to operate the magnetic unit by adding more electrons or by applying an electric field (ε). [25]. To demonstrate this idea, we included a positive ε in the self-consistent calculations and obtained the ε-dependence of MAE for Pt-Ir@SV. The presence of an electric field toward the graphene sheet noticeably reduces the magnetic moment and MAE of Pt-Ir@SV, as shown in Fig. 5c. For instance, its MAE becomes less than 10 meV at ε > 0.8 V/Å, easy for the writing operation. As shown in Fig. 5d, the main effect of the electric field is to induce charge rearrangement from $d_{z^2}^o$ orbital to $d_{xy/x^2-y^2}^u$ within the Pt atom. As a result of the additional occupation of the $d_{xy/x^2-y^2}^u$ states in the minority spin channel, their contributions to $M_S$ and MAE(dd) are reduced. In contrast, the actual Fermi level of Os-Ru@NDV lies in a plateau of its MAE($E_F - E_F^0$) curve hence MAE of Os-Ru@NDV does not change much with ε up to ±1.0 V/Å.

So far the MAEs were calculated by using the torque method that treats the SOC term in the Hamiltonian as a perturbation. [26,27] To check the reliability of these results for 5d systems, we also treated SOC self-consistently, with the non-collinear mode of the VASP. In principle, both spin and orbital magnetic moments of heavy atoms should depend on the orientation of magnetization and the easy axis is usually along the direction that maximizes spin and orbital moments. [8,9,28] As listed in Table III, the spin moment of Pt-Ir@SV changes from 0.60 μ$_B$ to 0.44 μ$_B$ and its orbital magnetic moment changes from 0.96 μ$_B$ to 0.34 μ$_B$, as the magnetization is switched from the easy axis to the hard axis. Large changes of spin (from 1.87 μ$_B$ to 1.28 μ$_B$) and orbital moments (from 1.29 μ$_B$ to 0.21 μ$_B$) are also observed for Os-Ru@NDV, as expected from its huge positive MAE. Self-consistent calculations give MAEs of 73 meV for Pt-Ir@SV and 54 meV for Os-Ru@NDV. These values are rather close to the corresponding values from the perturbative torque method, indicating the applicability of both approaches for the determination of MAEs. In particular, the torque approach is more efficient and practical for studies of large systems and also for broad searches as in the present work.

Strandberg et al and Xiao et al suggested that Co-Co, Rh-Rh and Ru-Ru dimers may have giant MAEs on graphene or benzene, but Błoński et al questioned their results. [7,8,10] We also calculated the MAEs of Co-Co, Rh-Rh and Ru-Ru dimers with both freestanding geometry and A-B@defect geometry as shown in Fig. 1. Our results indicate that the MAE of a free Co-Co dimer is only 5.2 meV, close to the result of Błoński et al (7.1 meV) but much less than those of Strandberg et al (28 meV) and Xiao et al (50 meV). Our MAE for the freestanding Rh-Rh dimer is 35 meV, also smaller than 52 meV and 104 meV obtained by Strandberg et al and Xiao et al, respectively. We found that only Rh-Rh@SV has a giant MAE of 60 meV, while the MAEs of other five cases are less than 6 meV. Moreover, the energy barrier for Rh-Rh dissociation along the pathway 4 as shown in Fig. 3a is only 0.4 eV, so the vertical geometry of Rh-Rh@SV is unstable at room temperature. Therefore, we see no potential to use Co, Rh and Ru dimers for applications.

Table III. Total and atom-resolved spin and orbital magnetic moments (in $\mu_B$) of Pt-Ir@SV and Os-Ru@NDV with the direction of magnetization along the vertical (⊥) or in-plane (↔) axis.

|  |  | $M_S$ | $M_S$ (A) | $M_S$ (B) | $M_L$ (A) | $M_L$ (B) |
|---|---|---|---|---|---|---|
| Pt-Ir@SV | ⊥ | 0.60 | 0.52 | 0.04 | 0.90 | 0.06 |
|  | ↔ | 0.44 | 0.42 | 0.01 | 0.33 | 0.01 |
| Os-Ru@NDV | ⊥ | 1.87 | 1.49 | 0.08 | 1.15 | 0.14 |
|  | ↔ | 1.28 | 1.03 | 0.03 | 0.12 | 0.09 |

Finally, we want to discuss the feasibility of fabricating the vertical Pt-Ir@SV and Os-Ru@NDV structures on graphene. We noted that embedding transition metal atoms in SVs and NDVs has been reported recently, [17,29,30] and the binding energies for Ir to SV and Ru to NDV of graphene are very large, as seen in table II. Therefore, there should be possible to first trap the B-type atoms (Ir and Ru) on SV and NDV. One extra step to obtain the Pt-Ir@SV and Os-Ru@NDV structures is the deposition of the A-type (Pt and Os) atoms. Pt and Os atoms are reasonably mobile on the perfect graphene, and Ir@SV and Ru@NDV act as attractive centers for them as indicated in Fig. 3. Therefore, the formation of vertical Pt-Ir and Os-Ru dimers should be straightforward in a well-controlled lab condition. The actual application of our prediction requires more fundamental and engineering studies, including finding appropriate protecting materials and fabrication conditions. To show the potential use of our systems, we estimate that an area density of 190 TB/in$^2$ is conceivable for magnetic recording and quantum computing operations, providing that the 8×8 graphene cells can be fabricated as the smallest magnetic units.

In summary, we performed systematic first-principles calculations to search for graphene-based nanostructures with giant magnetic anisotropy. Based on the criteria of MAE > 30 meV and high structural stability, we propose two structures with either Pt-Ir or Os-Ru dimer on the defected graphene as promising candidates for room temperature applications. Rigid band model analysis further indicates that the MAE of Pt-Ir@SV can be conveniently tuned by applying a moderate electric field. The design principle and the physical insights established here should be useful for the technological development of robust nanomagnetic units.

**Supporting Information**

A table of total and local spin moments of A-B@SV and A-B@NDV for all the transition-metal dimers. Figures for spin densities, PDOS, and effect of the Hubbard U for Ir-Pt@SV and Os-Ru@NDV. This material is available free of charge via the Internet at http://pubs.acs.org.

**Acknowledgements**

Work was supported by DOE-BES (Grant No: DE-FG02-05ER46237) and by NERSC for computing time.

Supporting Information for

# Giant magnetic anisotropy of transition-metal dimers on defected graphene

Jun Hu and Ruqian Wu

Department of Physics and Astronomy, University of California, Irvine, California 92697-4575, USA

**Table S1.** Total and local spin moments ($M_S$, in $\mu_B$) of A-B@SV and A-B@NDV without the spin-orbit coupling (SOC). The local spin moments are counted only in the Wigner-Seitz radii of the transition metal atoms. The magnetic anisotropy energies (MAE, in meV) of some selective cases from self-consistent total energy calculations with SOC are listed, with the MAEs from the torque method in the parentheses for comparison.

| A | B | A-B@SV | | | | A-B@NDV | | | |
|---|---|---|---|---|---|---|---|---|---|
|   |   | $M_S$ | $M_S$ (A) | $M_S$ (B) | MAE | $M_S$ | $M_S$ (A) | $M_S$ (B) | MAE |
| Os | Fe | 3.85 | 2.53 | 0.94 |  | 4.02 | 2.82 | 0.91 |  |
|    | Ru | 3.47 | 2.67 | 0.34 |  | 1.84 | 1.46 | 0.02 | 54 (63) |
|    | Os | 3.49 | 2.54 | 0.44 |  | 1.56 | 1.23 | 0.03 |  |
|    | Co | 2.99 | 2.58 | 0.05 |  | 4.05 | 3.17 | 0.62 |  |
|    | Rh | 2.99 | 2.59 | 0.00 |  | 3.52 | 2.97 | 0.29 |  |
|    | Ir | 2.87 | 2.46 | 0.01 |  | 3.35 | 2.75 | 0.28 |  |
| Ir | Fe | 2.75 | 1.49 | 1.07 |  | 3.02 | 1.77 | 1.15 |  |
|    | Ru | 2.37 | 1.64 | 0.49 |  | 3.00 | 1.85 | 0.75 |  |
|    | Os | 2.22 | 1.47 | 0.52 |  | 0.00 | 0.00 | 0.00 |  |
|    | Co | 1.87 | 1.65 | 0.21 |  | 3.26 | 2.17 | 0.88 |  |
|    | Rh | 1.86 | 1.72 | 0.10 |  | 2.05 | 1.75 | 0.20 |  |
|    | Ir | 1.75 | 1.54 | 0.14 |  | 2.11 | 1.66 | 0.30 |  |
| Pt | Fe | 1.90 | 0.57 | 1.35 |  | 2.03 | 0.70 | 1.36 |  |
|    | Ru | 1.57 | 0.68 | 0.69 |  | 1.95 | 0.72 | 0.82 |  |
|    | Os | 1.69 | 0.79 | 0.59 |  | 0.00 | 0.00 | 0.00 |  |
|    | Co | 0.61 | 0.51 | 0.04 |  | 1.01 | 0.62 | 0.52 |  |
|    | Rh | 0.63 | 0.56 | 0.02 |  | 0.98 | 0.63 | 0.28 |  |
|    | Ir | 0.60 | 0.48 | 0.05 | 73 (85) | 1.00 | 0.55 | 0.34 |  |

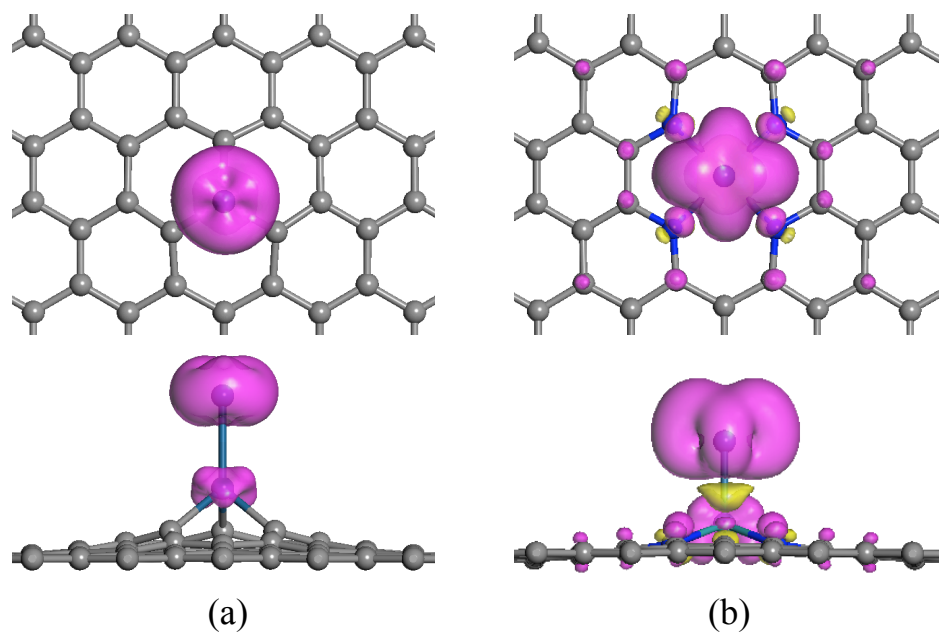

(a)    (b)

**Figure S1.** Isosurfaces of the spin density (at 0.01 e/Å$^3$) of (a) Ir-Pt@SV and (b) Os-Ru@NDV. The gray and blue balls stand for the C and N atoms, respectively. The transition metal atoms are enclosed in the isosurfaces. The pink and yellow isosurfaces show the positive and negative spin polarizations, respectively.

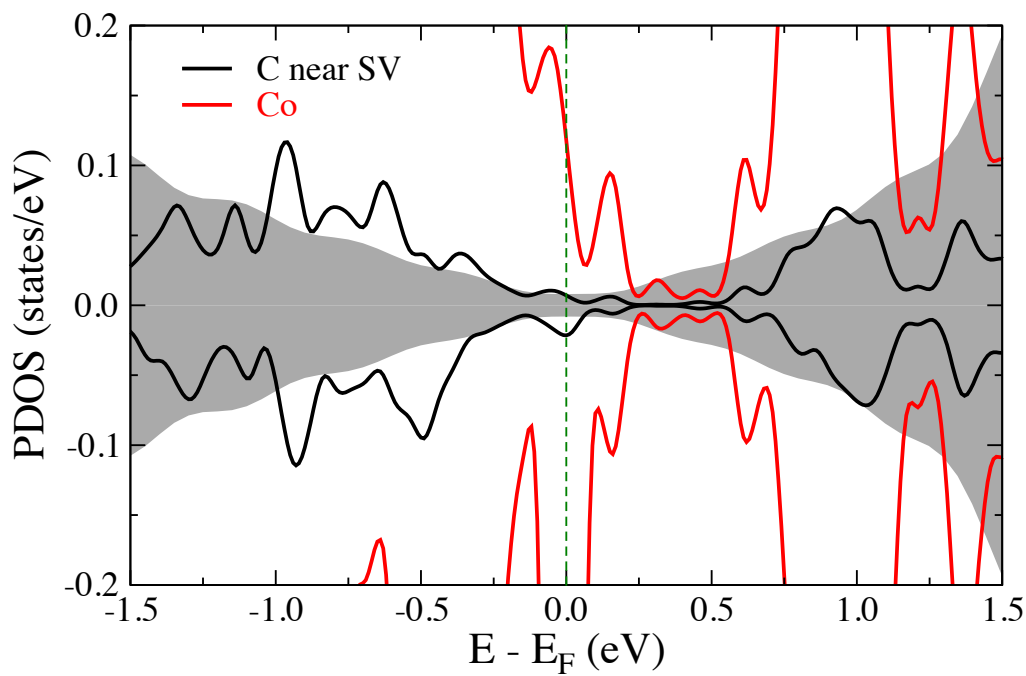

**Figure S2.** Projected density of states (PDOS) of Ir atom and nearby C atom of Pt-

Co@SV. The shadow area is the PDOS of C atom in pristine graphene.

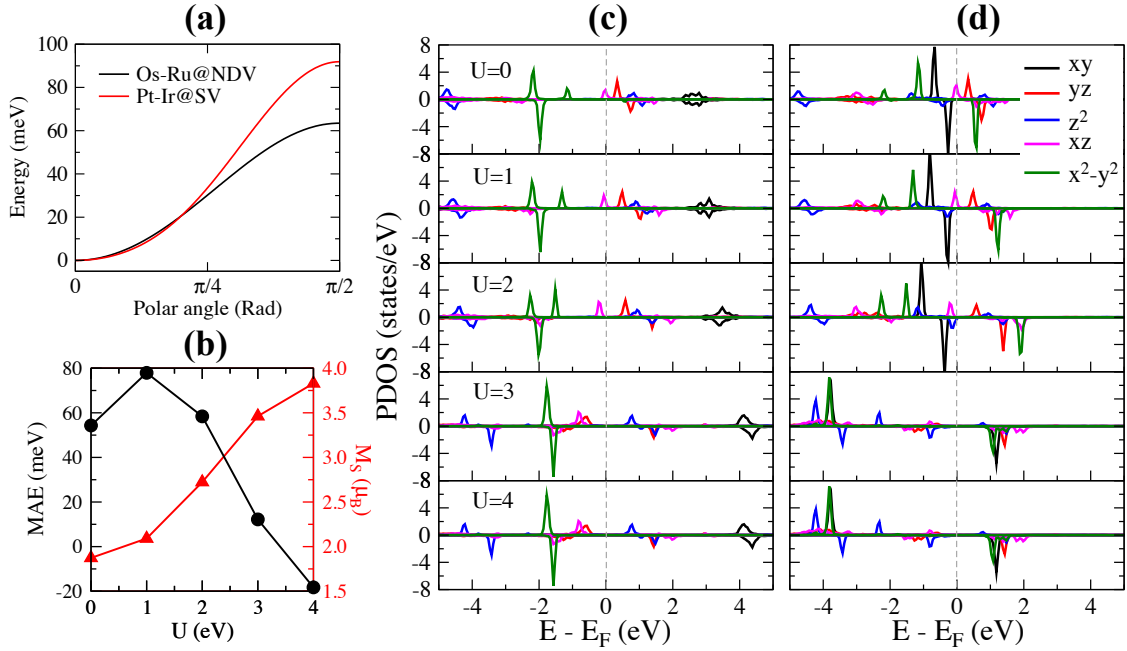

**Figure S3.** (a) The total energies of Pt-Ir@SV and Os-Ru@NDV as a function of the polar angle of the spin moment calculated by the torque method [S1]. (b) The MAEs and $M_S$ of Os-Ru@NDV as functions of Hubbard U from self-consistent calculations. (c) and (d) Projected density of states (PDOS) of Ru and Os, respectively.

**References**

[S1]. Wang, X. D.; Wu, R. Q.; Wang, D. S.; Freeman, A. J. *Phys. Rev. B* **1996**, 54, 61.